\documentclass[a4paper,fleqn,usenatbib]{mnras}

\usepackage{newtxtext,newtxmath}

\usepackage[T1]{fontenc}
\usepackage{ae,aecompl}

\usepackage{graphicx}	
\usepackage{amsmath}	
\usepackage{amssymb}	

\usepackage{stmaryrd}

\numberwithin{equation}{section}

\title[$\mathtt{FastChem}$]{$\mathtt{FastChem}$: A computer program for efficient complex chemical equilibrium calculations in the neutral/ionized gas phase with applications to stellar and planetary atmospheres}

\author[J. W. Stock et al.]{
Joachim W. Stock,$^{1}$\thanks{E-mail: joachimstock14@gmail.com}
Daniel Kitzmann,$^{2}$\thanks{E-mail: daniel.kitzmann@csh.unibe.ch}
A. Beate C. Patzer$^{3}$
and Erwin Sedlmayr$^{3}$
\\
$^{1}$Department of Chemistry and Environmental Science, Medgar Evers College-City University of New York, \\1650 Bedford Avenue, Brooklyn, NY 11235, United States\\
$^{2}$Center for Space and Habitability, University of Bern, Gesellschaftsstrasse. 6, 3012 Bern, Switzerland\\
$^{3}$Zenrum f\"ur Astronomie und Astrophysik (ZAA), Technische Universit\"at Berlin (TUB), Hardenbergstr. 36, 10623 Berlin, Germany
}

\date{Accepted XXX. Received YYY; in original form ZZZ}

\pubyear{2018}

\begin{document}
\label{firstpage}
\pagerange{\pageref{firstpage}--\pageref{lastpage}}
\maketitle

\begin{abstract}
For the calculation of complex neutral/ionized gas phase chemical equilibria, we present a semi-analytical versatile and efficient computer program, called \texttt{FastChem}. The applied method is based on the solution of a system of coupled nonlinear (and linear) algebraic equations, namely the law of mass action and the element conservation equations including charge balance, in many variables. Specifically, the system of equations is decomposed into a set of coupled nonlinear equations in one variable each, which are solved analytically whenever feasible to reduce computation time. Notably, the electron density is determined by using the method of Nelder and Mead at low temperatures. The program is written in object-oriented C++ which makes it easy to couple the code with other programs, although a stand-alone version is provided. \texttt{FastChem} can be used in parallel or sequentially and is available under the GNU General Public License version 3 at \url{https://github.com/exoclime/FastChem} together with several sample applications. The code has been successfully validated against previous studies and its convergence behavior has been tested even for extreme physical parameter ranges down to $100\,\mathrm{K}$ and up to $1000\,\mathrm{bar}$.
\texttt{FastChem} converges stable and robust in even most demanding chemical situations, which posed sometimes extreme challenges for previous algorithms.
\end{abstract}

\begin{keywords}
astrochemistry -- methods: numerical -- planets and satellites: atmospheres -- stars: atmospheres
\end{keywords}



\section{Introduction}
The detailed knowledge of the gas phase chemical composition is of manifold importance in astrophysics and planetary science. 
For example, it impacts not only the hydrodynamic and thermodynamic structure of stellar and planetary atmospheres, but also influences the spectral appearance of the object of interest by affecting the related radiative transport coefficients of absorption, spontaneous and induced emission.
Moreover, atomic and molecular species are the elementary building blocks of solid state bodies ranging in size from dust particles to rocky planets.

Chemical equilibrium (CE) situations occur, if the chemical timescale is much shorter than the dynamical timescale and if photochemical and cosmic ray induced processes can be neglected.
Whether the CE approximation is valid for a particular system has to be checked beforehand.
In non-CE situations, the mathematical CE solution is often used to define initial and/or boundary conditions for non-CE models.
The CE solution furthermore can be used as a reference state for comparison.
CE models have been used for example to calculate the chemical composition of the atmospheres of cool stars \citep[e.g.][]{Rus34,Var66,Tsu73,Joh82} and in dust driven winds of AGB (asymptotic giant branch) stars \citep[e.g.][]{Gai84,Gai86,Gai87,Dom90,Win94,Fer01}.
More recently, CE models were applied to atmospheres of brown dwarfs \citep{Bur02,Mar02,Hel08mnras,Hel08apj} and extrasolar planets \citep{Mad11,Kat14,Mor15}.

The CE composition is thermodynamically determined by the minimum of the Gibbs free energy of the system \citep[see][for example]{Den55,Ari69}.
Due to the nonlinear dependence of the total Gibbs free energy on the number densities of the chemical species involved in combination with linear constraints (element conservation, including charge balance and the requirement for the number densities to be nonnegative), the determination of the CE composition is in general non-trivial and often computational highly demanding \citep[see standard textbooks by e.g.][]{Zeg70,Smi82}.
Therefore, it is essential to develop rapid, efficient and versatile computer algorithms for the computation of complex chemical equilibria in general and equilibrium solutions of astrophysical objects in particular.

Different classification schemes for such numerical algorithms were proposed by e.g. \cite{Joh67}, \cite{Zeg70} and \cite{Smi82}, each of them emphasizing different aspects, such as e.g. Gibbs free energy method versus equilibrium constant method, treatment of the element abundance constraints and equilibrium conditions, numerical techniques or considerations of the total number of independent variables in terms of stoichiometric and nonstoichiometric algorithms \citep{Smi82}. 
A convenient way of classification is to distinguish between stoichiometric and nonstoichiometric algorithms first and then by the numerical method (i.e. minimization methods or methods solving sets of nonlinear algebraic equations).
Stoichiometric algorithms generally use reaction extends as free variables \citep{Nap59,Nap60,Nap61,Vil59,Cru64,Smi68}.
Although these methods resulted in applications in e.g. chemical engineering and combustion chemistry \citep[e.g.][]{Won04}, they are rarely, if at all, used in astrophysics or planetary atmospheric science, where nonstoichiometric methods based on optimization techniques or the solution of nonlinear algebraic equations are preferably applied.

The most commen optimization techniques, are the methods known as the RAND method \citep{Whi57,Whi58} and method of element potentials \citep{Pow59}, which are for example the basis of the computer codes SOLGAS \citep{Eri71}, SOLGASMIX \citep{Eri75}, SOLGASMIX-PV \citep{Bes77}, ChemSage \citep{Eri90}, NASA-CEA \citep{Gor94,Mcb96}, STANJAN \citep{Rey86}, and more recently TECA \citep{Ven12} and TEA \citep{Ble16}.

\citet{Bri47} developed a general algorithm to find the CE solution by solving a set of nonlinear algebraic equations, that is the law of mass action in combination with the atom balance condition \citep[see also][]{Bri46,Kan50} by using the Newton-Raphson method \citep[e.g.][]{Deu04}\footnote{For a historical note, see \citep{Kol92}} amongst others.
A method, known as the (original) NASA method, not to be confused with the algorithm implemented in the NASA-CEA code,  was developed by \citet{Huf51}.
\citet{Zel60,Zel68} showed the computationally equivalence between Brinkley's method, the NASA method and the RAND method.
Therefore, the three methods together are sometimes called BNR method \citep{Smi82}.
A similar algorithm to Brinkley's method was applied by \citet{Rus34} to investigate the CE composition of stellar atmospheres.
The main difference to the method of Brinkley is that it is restricted to diatomic molecules only and rests on the hierarchical structure of the element abundances.
Finding the CE solution with help of the law of mass action in combination with the element conservation by employing the Newton-Raphson method has been refined by e.g. \citet{Tsu64}, \citet{Var66} and \citet{Gai84} in order to take larger molecules and more species into account.
Algorithms of that kind have been implemented in codes such as CONDOR \citep{Lod93} and GGChem \citep[][and references therein]{Woi18}.

In this article we describe the computational efficient and fast algorithm of the computer code called \texttt{FastChem}, which calculates the CE composition of the gas phase for given gas pressure, temperature and specified element distribution.
The algorithm is loosely based on the method described by \citet{Gai14}.
One major difference to \citet{Gai14} is for example our treatment of the electron as a chemical species, for which the charge conservation is solved in \texttt{FastChem} with the method of Nelder and Mead \citep{Nel65} at low temperatures.

The \texttt{FastChem} program code is written in object-oriented C++ and is especially designed to be easily coupled with other models.
The released source code, however, also includes a stand-alone version.
\texttt{FastChem} can be run either with double or long double floating-point precision.
Long double usually\footnote{Note that the actual accuracy of double vs. long double depends on the employed compiler and computer platform. In some cases (e.g. Visual C++ compiler on a Windows operating system), long double and double will provide the same accuracy.} offers a higher precision and allows \texttt{FastChem} to compute the gas phase composition at very low temperatures.
While calculations in long double precision normally require a longer computation time per iteration, the total run time until convergence is reached can still be shorter than that of calculations in double precision. That is because the higher accuracy of long double floating-point precision can result in less iteration steps to be required.
Written in an object-oriented way, several instances of \texttt{FastChem} can be used simultaneously within one model.
For example, a double precision version of \texttt{FastChem} can be run for high temperatures, whereas a long double precision instance can be run for low temperature at the same time.
This also allows \texttt{FastChem} to be used in parallel by employing e.g OpenMP (Open Multi-Processing) \citep{Ope97}, which greatly increases the computational speed.
The chemistry model can also be parallelized with MPI \citep{Mcs97} by creating a separate \texttt{FastChem} instance on each MPI node, for example.
The \texttt{FastChem} code is released as open-source under the GNU General Public License version 3 \citep{Gnu07}.
The source code, together with several examples demonstrating various possible applications of \texttt{FastChem}, is published at: \url{https://github.com/exoclime/FastChem}.

\section{Method}
Let $\mathcal{S}$ be the set of all species in the gas phase without the electron and $\mathcal{E}\subset\mathcal{S}$ be the set of all chemical elements taken into account in the particular model.
Furthermore, let $\mathcal{S}_0$ be the of all species and $\mathcal{E}_0\subset\mathcal{S}_0$ be the set of all elements with the electron included.

The number densities $n_i$ for all species $S_i\in\mathcal{S}_0$ composed of elements $E_j\in\mathcal{E}_0$ for a given total gas pressure $p_\mathrm{g}$ and a given temperature $T$, are determined in dissociative equilibrium
\begin{equation}
S_i \rightleftharpoons \nu_{i0} E_0 + \nu_{i1} E_1 + \nu_{i2} E_2 + \ldots + \nu_{ij} E_j + \ldots = \sum_{j\in\mathcal{E}_0}\nu_{ij}E_j
\label{eq:reaction}
\end{equation}
by the law of mass action
\begin{equation}
n_i=K_i\prod_{j\in\mathcal{E}_0}n_j^{\nu_{ij}}\ ,\qquad \forall i\in \mathcal{S}\setminus\mathcal{E}\ ,
\label{eq:mwg}
\end{equation}
which can be derived by minimizing the Gibbs free energy of the system \citep[see e.g.][]{Den55,Ari69} in combination with the equations for element, respectively, charge conservation
\begin{equation}
\epsilon_j n_\mathrm{<H>}=n_j+\sum_{i\in\mathcal{S}\setminus\mathcal{E}}\nu_{ij}n_i\ ,\qquad\forall j\in\mathcal{E}_0\ .
\label{eq:ee}
\end{equation}
$K_i$ denotes the mass action constant, $\epsilon_j$ the relative elemental abundance with respect to hydrogen, $\nu_{ij}$ the coefficients of the stoichiometric matrix and 
\begin{equation}
n_\mathrm{<H>} = n_\mathrm{H} + \sum_{i\in\mathcal{S}\setminus\mathcal{E}}\nu_{i\mathrm{H}}n_i
\end{equation}
the sum of all hydrogen nuclei per unit volume.
The index 0 denotes the electron by definition, i.e. $\epsilon_0=0$ owing to the charge neutrality.
All stoichiometric coefficients $\nu_{ij}$ are nonnegative integers, if $j\neq0$.
For positively charged species $\nu_{i0}$ is a negative integer number, for negatively charged species $\nu_{i0}$ is a positive integer.
Otherwise $\nu_{i0}$ is $0$.
Any meaningful solution of the problem requires that the number densities $n_i$ are nonnegative
\begin{equation}
n_i\geq0\ ,
\label{eq:nonneg}
\end{equation}
which poses an additional constraint for the mathematical solution $\left\lbrace n_0,\ldots,n_{\left|\mathcal{S}_0\right|}\right\rbrace$.
Eqs~(\ref{eq:mwg}) and (\ref{eq:ee}) form a system of coupled nonlinear algebraic equations.
In the next two subsections the input and output data are decribed followed by the outline of the algorithm in Section~\ref{ssec:fastchem}.

\subsection{Input data}
For the computation of the chemical equilibrium solution, the algorithm needs as input a list of chemical elements, molecules and/or ions, which can be cations and/or anions of atoms and/or molecules, respectively.
Furthermore, the chemical element composition is given by
\begin{equation}
\epsilon_j=10^{x_j-12.0}\qquad ,
\label{eq:xj}
\end{equation}
using here the convention of stellar atmospheric theory, i.e. $x_\mathrm{H}=12$.
In addition, the natural logarithm of the dimensionless mass action constant of species $S_i\in\mathcal{S}\setminus\mathcal{E}$
\begin{equation}
\ln\bar{K}_i(T) = - \frac{\Delta_\mathrm{r} G_i^\minuso(T)}{R\,T}
\label{eq:lnK}
\end{equation}
for all considered molecules and ions, where $\Delta_\mathrm{r} G_i^{\minuso}(T)$ is the Gibbs free energy of dissociation (Eq.~(\ref{eq:reaction})) and $R$ the universal gas constant.
Gibbs free energies of dissociation	$\Delta_\mathrm{r} G_i^{\minuso}(T)$ can be calculated via
\begin{equation}
\Delta_\mathrm{r} G_i^{\minuso}(T)=\Delta_\mathrm{f} G_i^{\minuso}(T)-\sum_{j\in\mathcal{E}_0}\nu_{ij}\Delta_\mathrm{f} G_j^{\minuso}(T)\ ,\qquad i\in\mathcal{S}\setminus\mathcal{E}
\label{eq:drG}
\end{equation}
with the Gibbs free energies of formation $\Delta_\mathrm{f} G_i^{\minuso}(T)$ of species $S_i$, adopted from thermochemical databases such as e.g. \citet{Cha98}.
The mass action constants can be interpolated lookup tables which might prove to be time and memory consuming.
Here, we prefer to fit the data with the expression
\begin{equation}
\ln\bar{K}_i(T) = \frac{a_0}{T} + a_1\,\ln T + b_0 + b_1\,T + b_2\,T^2
\label{eq:fit}
\end{equation}
which we derived by using the van't Hoff equation \citep{Van84,Atk14} and Kirchoff's law of thermochemistry \citep{Kir58,Atk14} in combination with the quadratic expansion in temperature of the heat capacity $C_p^\minuso(T)$.
Thermochemical data are fitted within a prescribed temperature range depending on the available data.
Applying polynomial fits outside the the temperature range can lead to improper results \citep{Bur84}, especially if higher orders are involved.
Higher order terms are avoided in Eq.~(\ref{eq:fit}) so that extrapolation is at least possible in a limited range \citep[see discussion by][and references therein]{Woi18}.
New chemical species can be easily added, if their mass action constants are available, or removed by simply modifying the list of species in the input file.
The program code can also be adapted quite easily to a user specified input data format.

If requested, the user can provide the relative atomic masses of the elements $A_{\mathrm{r},j}$ for the additional calculation of the mean relative molecular mass according to
\begin{equation}
\left\langle M_{\mathrm{r},i} \right\rangle  = \frac{1}{n_\mathrm{g}}\sum_{i\in\mathcal{S}}n_i\,M_{\mathrm{r},i}\simeq\frac{1}{n_\mathrm{g}}\sum_{j\in\mathcal{E}}\epsilon_j\,n_\mathrm{<H>}\,A_{\mathrm{r},j}\qquad,
\end{equation}
where $M_{\mathrm{r},i}$ is the relative molecular mass.

Finally, the user has to specify at least one pair of $(p_\mathrm{g},T)$-data, where $p_\mathrm{g}$ is the gas pressure. Alternatively and computationally faster, the user can choose to provide the total pressure of hydrogen nuclei
\begin{equation}
p_\mathrm{<H>} = n_\mathrm{<H>}\,k_\mathrm{B}\,T
\label{eq:perfectgas}
\end{equation}
instead of $p_\mathrm{g}$. The Boltzmann constant is denoted by $k_\mathrm{B}$.

\subsection{Output data}
The output data is given in a formatted file listing the number densities $n_i$ of all species $S_i\in\mathcal{S}$.
In four separate columns the gas pressure $p_\mathrm{g}$, the temperature $T$, the total gas density $n_\mathrm{g}$ and $n_\mathrm{<H>}$ are added in the output file for visualization of the results with scientific information graphics software.
If requested, the output file also contains the mean relative molecular mass based on the specified relative atomic masses.

Moreover, the $\mathtt{FastChem}$ program generates a monitor file, listing the total number of pressure iterations, the total number of chemistry iterations in the last pressure iteration step and information about convergence. We recommend to carefully examine this file after the calculation is completed.

\subsection{Gas pressure -- total hydrogen nuclei density conversion}
\label{ssec:pressure}
If $p_\mathrm{<H>}$ is provided as input parameter for the calculation, then $n_\mathrm{<H>}$ is easliy obtained by Eq.~(\ref{eq:perfectgas})
However, for most practical applications, a total gas pressure $p_\mathrm{g}$ provided by the user rather than $p_\mathrm{<H>}$, is used as input.
In this case, the gas pressure is converted into $n_\mathrm{<H>}$ by the following procedure.
In H-He-rich gas mixtures at moderate temperatures $n_\mathrm{<H>}$ can be procproximated by
\begin{equation}
n_\mathrm{<H>} \approx n_\mathrm{H} + 2\,n_\mathrm{H_2}\ ,
\label{eq:eeh}
\end{equation}
which can be used to estimate the number density of He atoms
\begin{equation}
\epsilon_\mathrm{He} n_\mathrm{<H>}\approx n_\mathrm{He}\ .
\label{eq:eehe}
\end{equation}
Thereby, the contributions of other molecules, bearing less abundant elements and ions are neglected in both equations (Eq.~(\ref{eq:eeh}) and Eq.~(\ref{eq:eehe})).
The total gas number density is then simply obtained by
\begin{equation}
n_\mathrm{g}\approx n_\mathrm{H} + n_\mathrm{H_2} + n_\mathrm{He}\ ,
\label{eq:ngapprox}
\end{equation}
which yield a relation between $n_\mathrm{<H>}$ and the total gas density $n_\mathrm{g}$
\begin{equation}
n_\mathrm{<H>}\approx\frac{1}{2\,b^2 K_\mathrm{H_2}}\left( a + 4\,b\,K_\mathrm{H_2} n_\mathrm{g}-\sqrt{a^2+4\,b\,K_\mathrm{H_2} n_\mathrm{g}}\right) 
\end{equation}
after eliminating $n_\mathrm{H}$, $n_\mathrm{H_2}$, $n_\mathrm{He}$ by using Eq.~(\ref{eq:mwg}) for $i=\mathrm{H}_2$ and introducing the coefficients
\begin{equation}
a=1+\epsilon_\mathrm{He}
\end{equation}
and
\begin{equation}
b=1+2\,\epsilon_\mathrm{He}\ .
\end{equation}
This rough estimate of $n_\mathrm{<H>}$ is used as initial value $n_\mathrm{<H>}^{(0)}$ for the following iteration scheme.
After the number densities $n_i^{(k)}$, $k>0$ of all species $S_i\in\mathcal{S}_0$ are determined, the total number density
\begin{equation}
n_\mathrm{g}^{(k)}=\sum_{i\in\mathcal{S}_0}n_i^{(k)}
\end{equation}
is used instead of the approximation given by Eq.~(\ref{eq:ngapprox}).
If the calculated $n_\mathrm{g}^{(k)}$ is larger than $p_\mathrm{g}/(k_\mathrm{B}\,T)$ provided by the user, then we set
\begin{equation}
n_\mathrm{<H>}^{(k+1)} = (1-\lambda)\,n_\mathrm{<H>}^{(k)}\ .
\end{equation}
and
\begin{equation}
n_\mathrm{<H>}^{(k+1)} = \frac{1}{(1-\lambda)} n_\mathrm{<H>}^{(k)}
\end{equation}
otherwise, where $\lambda\in(0,1)$ is a damping parameter.
To avoid oscillations we reduce $\lambda$, if the expression $\left(p_\mathrm{g}/(k_\mathrm{B}\,T)-n_\mathrm{g}^{(k)}\right)$ changes sign between two iteration steps. 
The interation is continued until the convergence criterion
\begin{equation}
\left|\frac{p_\mathrm{g}}{k_\mathrm{B}\,T}-n_\mathrm{g}^{(k)}\right| < \delta \left|\frac{p_\mathrm{g}}{k_\mathrm{B}\,T}\right|\ , \qquad \delta>0
\end{equation}
is fulfilled.

\subsection{The $\mathtt{FastChem}$ algorithm}
\label{ssec:fastchem}
\subsubsection{Preconditioning}
The $\mathtt{FastChem}$ algorithm follows roughly the idea of the method presented by \citet{Gai14}.
That is, instead of solving the equations Eq.~(\ref{eq:mwg}) and Eq.~(\ref{eq:ee}) simultaneously, e.g. with a Newton-Raphson method in higher dimensions, the equation system is decomposed into a set of equations, each of them in one variable $n_j$, $j\in\mathcal{E}$.
Therefore, we rewrite the element conservation Eq.~(\ref{eq:ee}) with help of the law of mass action Eq.~(\ref{eq:mwg}) as 
\begin{equation}
\epsilon_j n_{<\mathrm{H}>}=n_j+\sum_{k=1}^{N_j}k\,n_j^k\underset{\epsilon_i=\epsilon_j}{\underset{\nu_{i j}=k}{\sum_{i\in \mathcal{S}\setminus\mathcal{E}}}}K_i \underset{l\neq j}{\prod_{l \in \mathcal{E}_0}} n_l^{\nu_{i l}}+n_{j,\mathrm{min}}\ ,\qquad j\in\mathcal{E}\ ,
\label{eq:basiceq}
\end{equation}
reducing the number of variables from $\left|\mathcal{S}_0\right|$ to $\left|\mathcal{E}_0\right|$ and solve Eq.~(\ref{eq:basiceq}) element by element, where
\begin{equation}
n_{j,\mathrm{min}}=\underset{\epsilon_i<\epsilon_j}{\sum_{i\in\mathcal{S}\setminus\mathcal{E}}}\nu_{ij}n_i\ ,\qquad j\in\mathcal{E}
\label{eq:nmin}
\end{equation}
is the total number density of all species build of elements less abundant than element $j$,
\begin{equation}
\epsilon_i=\min_{j\in\mathcal{E}}\left\lbrace \left. \epsilon_j\right|\nu_{ij}\neq 0\right\rbrace\ ,\qquad i\in\mathcal{S}\setminus\mathcal{E} 
\end{equation}
and
\begin{equation}
N_j=\max_{i\in\mathcal{S}\setminus\mathcal{E}}\left\lbrace\left. \nu_{ij}\right| \epsilon_i=\epsilon_j\right\rbrace\ ,\qquad j\in\mathcal{E}\ . 
\end{equation}
Solving Eq.~(\ref{eq:basiceq}) essentially reduces to the problem of finding the root of the polynomial 
\begin{equation}
P_j(n_j):=\sum_{k=0}^{N_j}A_{jk} n_j^k
\end{equation}
of degree $N_j$ for all $j\in\mathcal{E}$ with the coefficients
\begin{align}
A_{j0}&=-\epsilon_j n_{<\mathrm{H}>}+n_{j,\mathrm{min}}\\
A_{j1}&=1+\underset{\epsilon_i=\epsilon_j}{\underset{\nu_{i j}=1}{\sum_{i\in \mathcal{S}\setminus\mathcal{E}}}} K_i \underset{l\neq j}{\prod_{l \in \mathcal{E}_0}} n_l^{\nu_{i l}}\label{eq:coeffAj1}\\
A_{jk}&=k\underset{\epsilon_i=\epsilon_j}{\underset{\nu_{i j}=k}{\sum_{i\in \mathcal{S}\setminus\mathcal{E}}}} K_i \underset{l\neq j}{\prod_{l \in \mathcal{E}_0}} n_l^{\nu_{i l}},\qquad k\geq 2\ .\label{eq:coeffAjk}
\end{align}
The polynomials $P_j(n_j)$ and if needed their derivatives
\begin{equation}
P_j^\prime(n_j)=\sum_{k=1}^{N_j}k A_{jk} n_j^{k-1}
\end{equation}
are evaluated by employing Horner's rule \citep{Hor19}.
For the evaluation of the products in Eq.~(\ref{eq:coeffAj1}) and Eq.~(\ref{eq:coeffAjk}) see Appendix~\ref{sec:product}.

\subsubsection{Computational procedure}
To determine the CE composition, $\mathtt{FastChem}$ firstly sorts the elements according to their abundance in descending order using an adapted version of the $\mathtt{piksr2}$ algorithm \citep{Pre92}.
Afterwards, the set of equations Eq.~(\ref{eq:basiceq}) is solved iteratively, starting with the most abundant element, where $n_{j,\mathrm{min}}$ is employed as a correction term.
Since $n_{j,\mathrm{min}}$ is supposed to be relatively small, we set the initial values $n_{j,\mathrm{min}}^{(0)}=0$ for all elements except of carbon (C) in the carbon-rich case and oxygen (O) in the oxygen-rich case to account for the high bond energy of the CO molecule.
In the carbon-rich case we set $n_{\mathrm{C,min}}^{(0)}=\epsilon_\mathrm{O} n_\mathrm{<H>}$ and in the oxygen-rich case we set $n_{\mathrm{O,min}}^{(0)}=\epsilon_\mathrm{C} n_\mathrm{<H>}$.
The initial electron density $n_0^{(0)}$ is always set to a very small value.
The $\mathtt{FastChem}$ algorithm works as follows:
\begin{description}
	\item[\textbf{Step 1}]\hfill\\Initial values for the electron density $n_0^{(0)}$ and for the correction terms $n_{j,\mathrm{min}}^{(0)}$ are set and the logarithmic mass action constants $\ln K_i$ are calculated for a given temperature $T$.
	\item[\textbf{Step 2}]\hfill\\The number densities for all atomic species $n_j\ (j\in\mathcal{E}$) are calculated via Eq.~(\ref{eq:basiceq}) (or Eq.~(\ref{eq:basiceqalt}) if necessary) in descending order, starting with the most abundant element.
	\item[\textbf{Step 3}]\hfill\\The results are used to calculate the number densities of the molecular species $n_i\ (i\in\mathcal{S}\setminus\mathcal{E})$ via the law of mass action Eq.~(\ref{eq:mwg}).
	\item[\textbf{Step 4}]\hfill\\$n_{j,\mathrm{min}}$ is updated.
	\item[\textbf{Step 5}]\hfill\\The electron density $n_0$ is calculated (see Section~\ref{sssec:electron}).
\end{description}
Steps 2 to 5 are repeated until the convergence criterion
\begin{equation}
\left|n_i^{(k)}-n_i^{(k-1)} \right| > \delta \left|n_i^{(k)} \right|\ ,\qquad\delta>0
\label{eq:convergence} 
\end{equation}
is met for all species $i\in \mathcal{S}$.

The procedure depends on whether $A_{j0}$ is a strictly negative number or not.
If $n_j$ is the solution of Eq~(\ref{eq:basiceq}), it is quite evident that $A_{j N_j}>0$, $A_{j k}\geq 0$ for $0<k<N_j$ and $A_{j 0}< 0$.
However, during the iteration it might happen that $A_{j0}$ becomes positive.
This situation can occur, if there are three elements, say X, Y and Z, with $\epsilon_\mathrm{X}\gtrapprox\epsilon_\mathrm{Y}$ and  $\epsilon_\mathrm{X}\gtrapprox\epsilon_\mathrm{Z}$ forming two molecules XY and XZ with large mass action constants $K_\mathrm{XY}$ and $K_\mathrm{XZ}$.
The molecules XY and XZ, competing for the element X, lead then to a large correction $n_\mathrm{X,min}$ which can exceed $\epsilon_\mathrm{X}n_\mathrm{<H>}$.
 
We first consider the case $A_{j0}<0$.
It can be easily shown that there exists a unique solution of Eq~(\ref{eq:basiceq}) which suffice the condition Eq.~(\ref{eq:nonneg}).
If $N_j$ is less than three, there are simple analytical expressions which provides the (temporary) solution for the number density $n_j$, namely
\begin{equation}
n_j=-A_{j0}
\begin{cases}
1\ ,\qquad&\mathrm{if}\ N_j=0\\
1/A_{j1}\ ,\qquad&\mathrm{if}\ N_j=1\\
2/\left( A_{j1}+\sqrt{A_{j1}^2-4 A_{j2} A_{j0}}\right) \ ,\qquad&\mathrm{if}\ N_j=2
\end{cases}\ .
\end{equation}
Otherwise, we employ the classical ordinary Newton-Raphson method \citep{Deu04} in one dimension
\begin{equation}
n_j^{(\mu+1)}=n_j^{(\mu)}-\frac{P_j^{(\mu)}}{P_j^{\prime(\mu)}}
\end{equation}
to obtain $n_j$.
Since $P_j(n_j)$ is two times continuously differentiable, convex in the interval $\left[0,\epsilon_j n_\mathrm{<H>}\right]$, $P_j(0)<0$ and $P_j(\epsilon_j n_\mathrm{<H>})>0$, there exists only one unique root $n_j^*$ and hence the Newton-Raphson method converges monotonously against the mathematical solution for suitable initial values, i.e. $n_j^{(0)}>n_j^*$.
Therefore, no computational costly damping strategy is required.
To guarantee convergence of the Newton-Raphson method, we set the initial value $n_j^{(0)}=\epsilon_j n_\mathrm{<H>}$.

If $A_{j0}\geq0$, then no real root of $P_j(n_j)$ exists, which suffice the condition $n_j>0$.
In this case, Eq.~(\ref{eq:basiceq}) in Step 2 is replaced by
\begin{equation}
\epsilon_j n_{<\mathrm{H}>}=n_j + \sum_{k=1}^{N_j}k\,n_j^k\underset{\nu_{i j}=k}{\sum_{i\in \mathcal{S}\setminus\mathcal{E}}}K_i \underset{l\neq j}{\prod_{l \in \mathcal{E}_0}} n_l^{\nu_{i l}}\ ,\qquad j\in\mathcal{E}
\label{eq:basiceqalt}
\end{equation}
for element $E_j$ for all remaining \texttt{FastChem} iterations.
Moreover $n_{j,\mathrm{min}}$ does not need to be updated for that element anymore.
If the procedure fails to converge within a prescribed number of iterations, \texttt{FastChem} starts again and tries to find the CE solution by solving Eq.~(\ref{eq:basiceqalt}) for all elements $E_j\in\mathcal{E}$.
In case this backup procedure fails, the computation is aborted with an error message.

The algorithm computes the CE composition $\left\lbrace n_0,\ldots,n_{\left|\mathcal{S}_0\right|}\right\rbrace$ for each pair $(p_\mathrm{g},T)$ or $(p_\mathrm{<H>},T)$ separately.
For large sets of $(p_\mathrm{g},T)$-pairs or $(p_\mathrm{<H>},T)$-pairs the computational speed might be increased by using the solutions of neighboring grid points as initial values $n_i^{(0)}$.
However, it might happen in such a case, that some of the in this way attributed values/initial values underestimate the number densities $n_j$, $j\in\mathcal{E}$, of the CE solution, which can impair the convergence behavior of the Newton solver.
Therefore, this approach is not implemented in the $\mathtt{FastChem}$ code so far.

\subsubsection{Determination of the electron density}
\label{sssec:electron}
For the electron species Eq.~(\ref{eq:basiceq}) becomes a homogeneous equation.
Although, at least for systems including only ions of ionization degree one ($\left|\nu_{i0}\right|=1$), an analytic solution can be derived (see Appendix~\ref{sec:edensity}), we follow a different approach here.
If there are sufficient free electrons available, the electron density $n_0$ can be calculated from the sum of the ion densities
\begin{equation}
n_0=-\sum_{i\in\mathcal{S}}\nu_{i0}n_i\ .
\end{equation}
Note that $\nu_{i0}>0$ for anions and $\nu_{i0}<0$ for cations.
That approach, however becomes problematic at low temperatures and/or high pressures due to cancelation of leading digits.
Therefore, if $n_0 < 0.9\,n^+$, we employ the method of Nelder and Mead \citep{Nel65,Lag97}, also known as the downhill simplex method \citep{Pre92}.
The method of Nelder and Mead is designed to find minima of multivariable functions $f:\mathbb{R}^n\longrightarrow\mathbb{R}$ and is especially successful for low dimensions $n\in\mathbb{N}$.
Furthermore, no evaluation of the derivative or the Jacobi matrix of $f$ is necessary.
We have reduced the method to one dimension.
To find the mathematical solution of the eletron density $n_0$, we introduce the objective function
\begin{align}
f(y_0)\,&:=\left|e^{y_0}+\sum_{i\in\mathcal{S}}\nu_{i0}n_i\right|\\
&=\left|e^{y_0}+\sum_{i\in\mathcal{S}}\nu_{i0}\exp\left\lbrace\ln K_i + \nu_{i0}y_0+\sum_{j\in\mathcal{E}}\nu_{ij}\ln n_j \right\rbrace \right|
\end{align}
with $y_0=\ln n_0$.
In the second part of the equation, the law of mass action Eq.~(\ref{eq:mwg}) is used.
To dampen numerical oscillations we modify the the electron density at the iteration step $\mu$ according to
\begin{equation}
n_0^{(\mu)}=\sqrt{n_0^{(\mu)}n_0^{(\mu-1)}}\ .
\end{equation}

\section{Results and Discussion}
\subsection{Test calculations}
In order to validate our code, we compare the results of $\mathtt{FastChem}$ with the pure gas phase results presented by \citet{Sha90}, who in turn compared their results with the ones presented by \citet{Tar87}.
In our test calculation, we adopt a fixed total hydrogen nuclei pressure $p_\mathrm{<H>}= 1000\,\mathrm{dyn}\,\mathrm{cm}^{-2}$ and the solar photospheric element abundances determined by \citet{Asp09} for elements more abundant than Vanadium (V) (see Table~\ref{table:elementabundances}).
\begin{table}
	\caption{Solar photospheric element abundances $x_j$ according to \citet{Asp09} for all elements included in the test example.} 
	\label{table:elementabundances} 
	\centering 
	\begin{tabular}{l l r | l l r} 
		\hline\hline 
		\multicolumn{2}{c}{Element} & $x_j$ & \multicolumn{2}{c}{Element} & $x_j$ \\ 
		\hline 
		H  & Hydrogen  & 12.00 & Na & Sodium     & 6.24\\
		He & Helium    & 10.93 & Ni & Nickel     & 6.22\\ 
		O  & Oxygen    &  8.69 & Cr & Chromium   & 5.64\\
		C  & Carbon    &  8.43 & Cl & Chlorine   & 5.50\\
		Ne & Neon      &  7.93 & Mn & Manganese  & 5.43\\
		N  & Nitrogen  &  7.83 & P  & Phosphorus & 5.41\\
		Mg & Magnesium &  7.60 & K  & Potassium  & 5.03\\
		Si & Silicon   &  7.51 & Co & Cobalt     & 4.99\\
		Fe & Iron      &  7.50 & Ti & Titanium   & 4.95\\
		S  & Sulfur    &  7.12 & F  & Fluorine   & 4.56\\
		Al & Aluminium &  6.45 & Zn & Zinc       & 4.56\\
		Ar & Argon     &  6.40 & Cu & Copper     & 4.19\\
		Ca & Calcium   &  6.34 & V  & Vanadium   & 3.93\\
		\hline 
	\end{tabular}
\end{table}
The set of species used in our test calculation encompasses all molecules and ions listed by \citet{Cha98} formed from the elements shown in Table~\ref{table:elementabundances}.
The list of species is complemented by molecules of potential astrophysical interest \citep{Tsu73}.
Therefore, in contrast to \citet{Sha90}, we use a slightly different set of species, which does not include the low abundant elements Scandium (Sc), Zirconium (Zr), Strontium (Sr), Bromine (Br) and Yttrium (Y).
However, our standard scenario includes the noble gases Helium (He), Neon (Ne) and Argon (Ar), the elements Cobalt (Co) and Zinc (Zn), ions, and numerous additional molecules, mostly chlorides and fluorides (see Table~\ref{table:species}). 
\begin{table*}
	\caption{List of all species included in the test model. Mass action constants $\ln\bar{K}_i$ are fitted according to Eq.~(\ref{eq:fit}) using thermochemical data from \citet{Cha98} unless indicated otherwise.} 
	\label{table:species} 
	\centering 
	\begin{tabular}{l p{16cm}} 
		\hline\hline 
		Element & Molecules, Ions \\ 
		\hline 
		H  & H$_2$, H$^+$, H$^-$, H$_2^+$, H$_2^-$\\
		He & He$^+$\\
		O  & HO, HO$_2$, H$_2$O, H$_2$O$_2$$^a$ ,O$_2$, O$_3$, HO$^+$, HO$^-$, H$_3$O$^+$, 
		O$^+$, O$^-$, O$_2^+$, O$_2^-$\\
		C  & CH, CHO, CH$_2$, CH$_2$O, CH$_3$, CH$_4$, CO, 
		CO$_2$, C$_2$, C$_2$H, C$_2$H$_2$, C$_2$H$_4$, C$_2$H$_4$O, C$_2$O, C$_3$, C$_3$H$^b$, 
		C$_3$O$_2$, C$_4$, C$_5$, C$^+$, C$^-$, CH$^+$, CH$^-$$^b$, CHO$^+$, 
		CO$_2^-$, C$_2^-$\\
		Ne & Ne$^+$\\
		N  & CHN, CHNO, CN, CNO, CNN, NCN, 
		C$_2$N, C$_2$N$_2$, C$_4$N$_2$, HN, HNO, c-HNO$_2$, t-HNO$_2$, 
		HNO$_3$, H$_2$N, H$_2$N$_2$, H$_3$N, H$_4$N$_2$, NO, NO$_2$, NO$_3$, 
		N$_2$, N$_2$O, N$_2$O$_3$, N$_2$O$_4$, N$_2$O$_5$, N$_3$, CN$^+$, CN$^-$, N$^+$, 
		N$^-$, NO$^+$, NO$_2^-$, N$_2^+$, N$_2^-$, N$_2$O$^+$\\
		Mg & HMg, HMgO, H$_2$MgO$_2$, MgN, MgO, Mg$_2$, 
		HMgO$^+$, Mg$^+$\\
		Si & CSi, CSi$_2$, C$_2$Si$^b$, C$_2$Si$_2$, HSi, H$_2$Si$^b$, H$_3$Si$^b$, 
		H$_4$Si, NSi, NSi$_2$, OSi, O$_2$Si, Si$_2$, Si$_3$, HSi$^+$, 
		Si$^+$, Si$^-$\\
		Fe & C$_5$FeO$_5$, FeH$_2$O$_2$, FeO, Fe$^+$, Fe$^-$\\
		S  & COS, CS, CS$_2$, FeS, HS$^c$, H$_2$O$_4$S, H$_2$S, 
		MgS, NS, OS, OS$_2$$^c$, O$_2$S, O$_3$S, SSi, S$_2$, S$_3$, 
		S$_4$, S$_5$, S$_6$, S$_7$, S$_8$, HS$^-$$^b$, S$^+$, S$^-$\\
		Al & AlH, AlHO(1), AlHO(2), AlHO$_2$, AlN, 
		AlO, AlO$_2$, AlS, Al$_2$, Al$_2$O, Al$_2$O$_2$, CAl, Al$^+$, 
		Al$^-$, AlHO$^+$, AlHO$^-$, AlO$^+$, AlO$^-$, AlO$_2^-$, 
		Al$_2$O$^+$, Al$_2$O$_2^+$\\
		Ar & Ar$^+$\\
		Ca & CaH$^b$, CaHO, CaH$_2$O$_2$, CaO, CaS, Ca$_2$, 
		Ca$^+$, CaHO$^+$\\
		Na & CNNa, C$_2$N$_2$Na$_2$, HNa, HNaO, H$_2$Na$_2$O$_2$, NaO, 
		Na$_2$, Na$_2$O$_4$S, HNaO$^+$, Na$^+$, Na$^-$, NaO$^-$\\
		Ni & C$_4$NiO$_4$, HNi$^b$, NiO$^b$, NiS, Ni$^+$, Ni$^-$\\
		Cr & C$_2$Cr$^b$, CrH$^b$, CrN, CrO, CrO$_2$, CrO$_3$, Cr$^+$, 
		Cr$^-$\\
		Cl & AlCl, AlClO, AlCl$_2$, AlCl$_3$, Al$_2$Cl$_6$, CCl, 
		CClN, CClO, CCl$_2$, CCl$_2$O, CCl$_3$, CCl$_4$, 
		CHCl, CHCl$_3$, CH$_2$Cl$_2$, CH$_3$Cl, C$_2$Cl$_2$, C$_2$Cl$_4$, 
		C$_2$Cl$_6$, C$_2$HCl, CaCl, CaCl$_2$, ClFe, ClH, ClHO, 
		ClH$_3$Si, ClMg, ClNO, ClNO$_2$, ClNa, ClNi, 
		ClO, ClO$_2$, ClO$_3$, ClS, ClS$_2$, ClSi, Cl$_2$, 
		Cl$_2$Fe, Cl$_2$H$_2$Si, Cl$_2$Mg, Cl$_2$Na$_2$, Cl$_2$Ni, ClOCl, 
		ClClO, ClO$_2$Cl, ClOClO, Cl$_2$O$_2$S, Cl$_2$S, 
		Cl$_2$Si, Cl$_3$Fe, Cl$_3$HSi, Cl$_3$Si, Cl$_4$Fe$_2$, Cl$_4$Mg$_2$, 
		Cl$_4$Si, Cl$_6$Fe$_2$, AlCl$^+$, AlCl$_2^+$, AlCl$_2^-$, Cl$^+$, 
		ClMg$^+$, ClS$^+$, Cl$^-$, Cl$_2$S$^+$\\
		Mn & HMn$^b$, MnO$^b$, MnS$^b$, Mn$^+$\\
		P  & CHP, CP, ClP, Cl$_3$OP, Cl$_3$P, Cl$_3$PS, 
		Cl$_5$P, HP, H$_2$P, H$_3$P, NP$^d$, OP$^d$, O$_2$P, O$_6$P$_4$$^d$, 
		O$_{10}$P$_4$, PS$^c$, P$_2$, P$_4$, P$_4$S$_3$, P$^+$, P$^-$\\
		K  & CKN, C$_2$K$_2$N$_2$, ClK, Cl$_2$K$_2$, HK, HKO, 
		H$_2$K$_2$O$_2$, KO, K$_2$, K$_2$O$_4$S, HKO$^+$, K$^+$, K$^-$, KO$^-$\\
		Co & ClCo, Cl$_2$Co, Cl$_3$Co, Cl$_4$Co$_2$, Co$^+$, Co$^-$\\
		Ti & C$_2$Ti$^b$, C$_4$Ti$^b$, ClOTi, ClTi, Cl$_2$OTi, Cl$_2$Ti, 
		Cl$_3$Ti, Cl$_4$Ti, NTi$^b$, OTi, O$_2$Ti, STi$^b$, Ti$^+$, Ti$^-$\\
		F  & AlClF, AlClF$_2$, AlCl$_2$F, AlF, AlFO, 
		AlF$_2$, AlF$_2$O, AlF$_3$, AlF$_4$Na, Al$_2$F$_6$, CClFO, 
		CClF$_3$, CCl$_2$F$_2$, CCl$_3$F, CF, CFN, CFO, CF$_2$, 
		CF$_2$O, CF$_3$, CF$_4$, CF$_4$O, CF$_8$S, CHF, CHFO, 
		CHF$_3$, CH$_2$ClF, CH$_2$F$_2$, CH$_3$F, C$_2$F$_2$, C$_2$F$_3$N, C$_2$F$_4$, 
		C$_2$F$_6$, C$_2$HF, CaF, CaF$_2$, ClF, ClFMg, ClFO$_2$S, 
		ClFO$_3$, ClF$_2$OP, ClF$_3$, ClF$_3$Si, ClF$_5$, ClF$_5$S, 
		CHClF$_2$, CHCl$_2$F, Cl$_2$FOP, Cl$_3$FSi, CoF$_2$, FFe, 
		FH, FHO, FHO$_3$S, FH$_3$Si, FK, FMg, FN, 
		FNO, FNO$_2$, FNO$_3$, FNa, FO, FOTi, OFO, 
		FOO, FP, FPS, FS, FSi, FTi, F$_2$, 
		F$_2$Fe, F$_2$H$_2$, F$_2$H$_2$Si, F$_2$K$_2$, F$_2$Mg, F$_2$N, c-F$_2$N$_2$, 
		t-F$_2$N$_2$, F$_2$Na$_2$, F$_2$O, F$_2$OS, F$_2$OSi, F$_2$OTi, F$_2$O$_2$, 
		F$_2$O$_2$S, F$_2$P, F$_2$S, F$_2$S$_2$(1), F$_2$S$_2$(2), F$_2$Si, F$_2$Ti, 
		F$_3$Fe, F$_3$HSi, F$_3$H$_3$, F$_3$N, F$_3$NO, F$_3$OP, F$_3$P, 
		F$_3$PS, F$_3$S, F$_3$Si, F$_3$Ti, F$_4$H$_4$, F$_4$Mg$_2$, F$_4$N$_2$, F$_4$S, 
		F$_4$Si, F$_4$Ti, F$_5$H$_5$, F$_5$P, F$_5$S, F$_6$H$_6$, F$_6$S, F$_7$H$_7$, 
		F$_{10}$S$_2$, AlClF$^+$, AlF$^+$, AlF$_2^+$, AlF$_2^-$, AlF$_2$O$^-$, 
		AlF$_4^-$, CF$^+$, CF$_2^+$, CF$_3^+$, F$^+$, F$^-$, FMg$^+$, FP$^+$, 
		FP$^-$, FS$^+$, FS$^-$, F$_2$K$^-$, F$_2$Mg$^+$, F$_2$Na$^-$, F$_2$P$^+$, 
		F$_2$P$^-$, F$_2$S$^+$, F$_2$S$^-$, F$_3$S$^+$, F$_3$S$^-$, F$_4$S$^+$, F$_4$S$^-$, 
		F$_5$S$^+$, F$_5$S$^-$, F$_6$S$^-$\\
		Zn & Zn$^+$, Zn$^-$\\
		Cu & ClCu, Cl$_3$Cu$_3$, CuF, CuF$_2$, CuH$^b$, CuO, 
		CuS$^b$, Cu$_2$, Cu$^+$, Cu$^-$\\
		V  & C$_2$V$^b$, C$_4$V$^b$, NV, OV, O$_2$V, V$^+$, V$^-$\\
		\hline 
		\multicolumn{2}{p{17cm}}{$^a$For H$_2$O$_2$, the data by \citet{Cha98} are only tabulated up to $T=2000\,\mathrm{K}$. Extrapolation of $\ln\bar{K}_\mathrm{H_2O_2}$ to higher temperatures yields good agreement with data obtained by \cite{Goo16}.}\\
		\multicolumn{2}{l}{$^b$ Thermochemical data by \citet{Tsu73} used to fit $\ln\bar{K}_i$.}\\
		\multicolumn{2}{l}{$^c$ Thermochemical data from \citet{Bar95} used to fit $\ln\bar{K}_i$.}\\
		\multicolumn{2}{l}{$^d$ Thermochemical data by \citet{Bur05} and \citet{Goo16} used to fit $\ln\bar{K}_i$.}\\
	\end{tabular}
\end{table*}
Logarithmic mass action constants $\ln\bar{K}_i$ were determined using
thermochemical data from \citet{Bar95}, \citet{Bur05}, \citet{Cha98}, \citet{Goo16} and \citet{Tsu73} as indicated in Table~\ref{table:species}.
\citet{Bar95} and \citet{Cha98} provide Gibbs free energy data $\Delta_\mathrm{f} G_i^\minuso(T)$ in tabulated form. From these data $\ln\bar{K}_i$ were calculated according to Eq.~(\ref{eq:lnK}) and Eq.~(\ref{eq:drG}). 
The obtained mass action constants $\ln\bar{K}_i$ were then subsequently fitted according to Eq~(\ref{eq:fit}).
\citet{Bur05}, \citet{Goo16} and \citet{Tsu73} provide thermochemical data in form of polynomials in temperature $T$ or reciprocal temperature $\theta=5040/T$, from which we obtained $\ln\bar{K}_i$ values, which we then fitted to derive the coefficients used in Eq~(\ref{eq:fit}).

\begin{figure}
	\resizebox{\hsize}{!}{\includegraphics{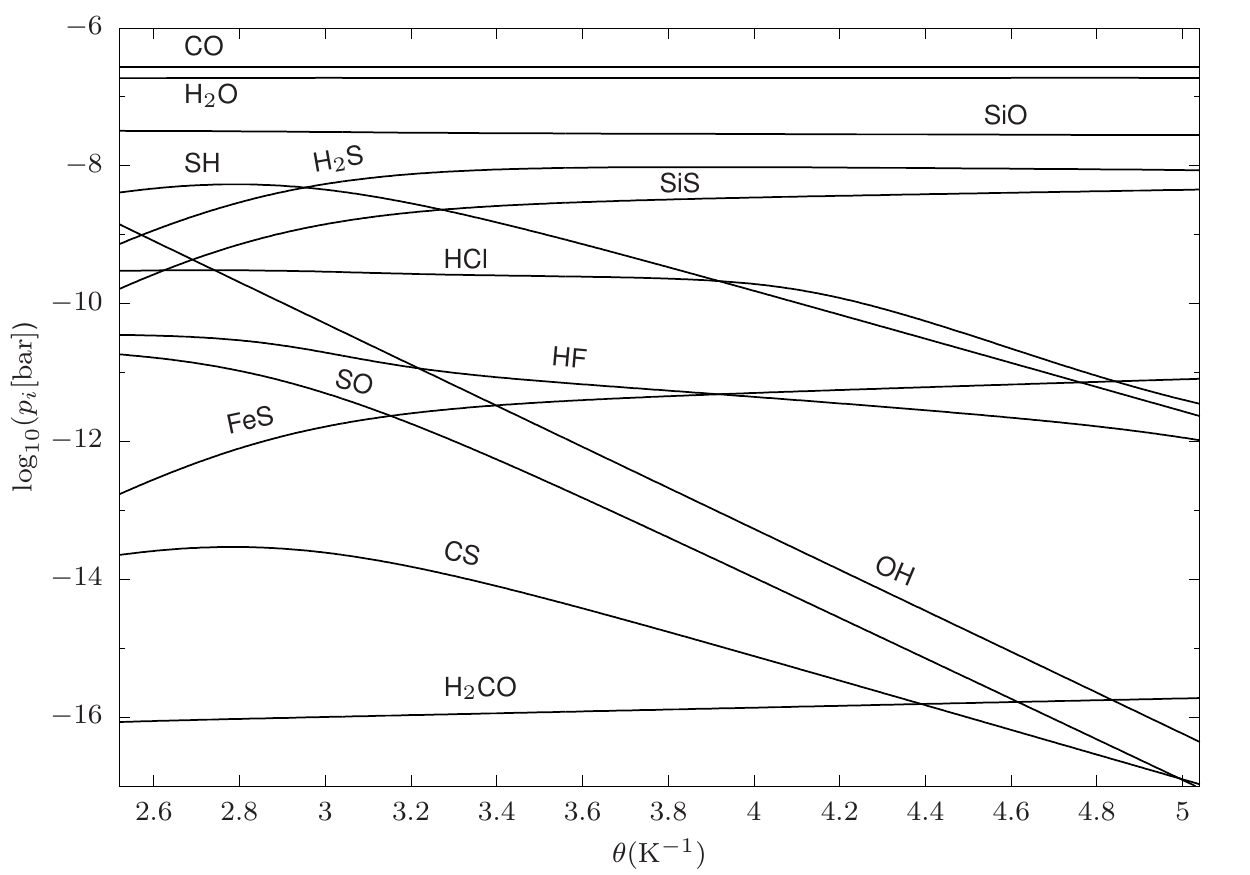}} 
	\caption{Logarithmic partial pressures of the molecular species CO, H$_2$O, SiO, SH, H$_2$S, OH, SiS, HCl, HF, SO, FeS, CS and H$_2$CO calculated with $\mathtt{FastChem}$. The total hydrogen nuclei pressure $p_\mathrm{<H>}$ equals $0.01\,\mathrm{bar}$ as in the calculations by \citet{Tar87} and \citet{Sha90}.} 
	\label{fig:sh1} 
\end{figure}

Figure~\ref{fig:sh1} shows the same species as presented by \citet{Sha90} and \citet{Tar87}, respectively.
In overall, we find good qualitative agreement with \cite{Tar87} and very good qualitative agreement with \citet{Sha90}.
Like \citet{Sha90}, we find that in contrast to the results of \citet{Tar87} H$_2$S to be more abundant than SiS and FeS more abundant than CS at $\theta=2.65263\,\mathrm{K}^{-1}$ ($T=1900\,\mathrm{K}$), where $\theta=5040/T$.

\begin{figure}
	\resizebox{\hsize}{!}{\includegraphics{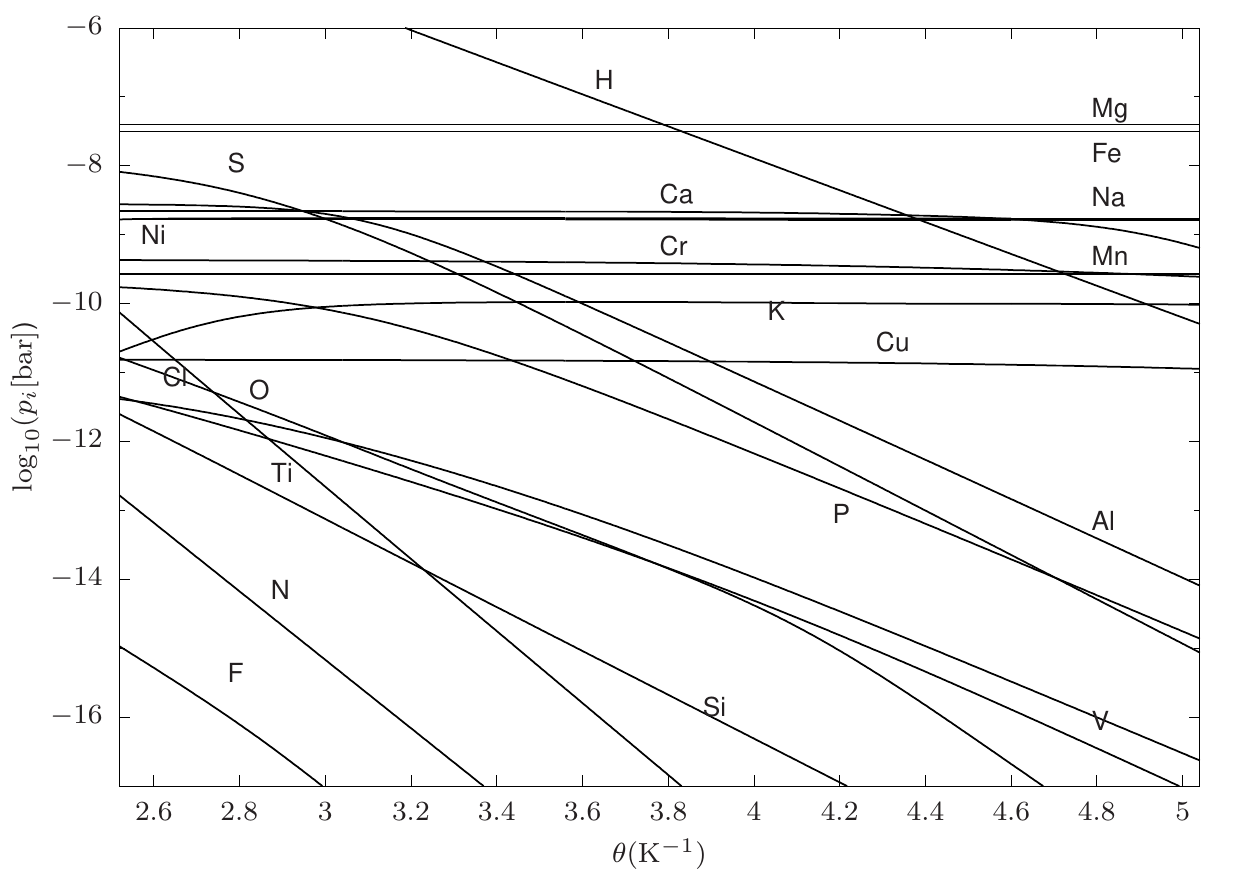}} 
	\caption{Logarithmic partial pressures of the the atomic species H, N, O, F, Na, Mg, Al, Si, P, S, Cl, K, Ca, Sc, Ti, V, Cr, Mn, Fe, Ni, Cu, Br, Sr and Zr.  The total hydrogen nuclei pressure $p_\mathrm{<H>}$ equals $0.01\,\mathrm{bar}$ as in the calculations by \citet{Tar87} and \citet{Sha90}.} 
	\label{fig:sh6} 
\end{figure}

\begin{figure}
	\resizebox{\hsize}{!}{\includegraphics{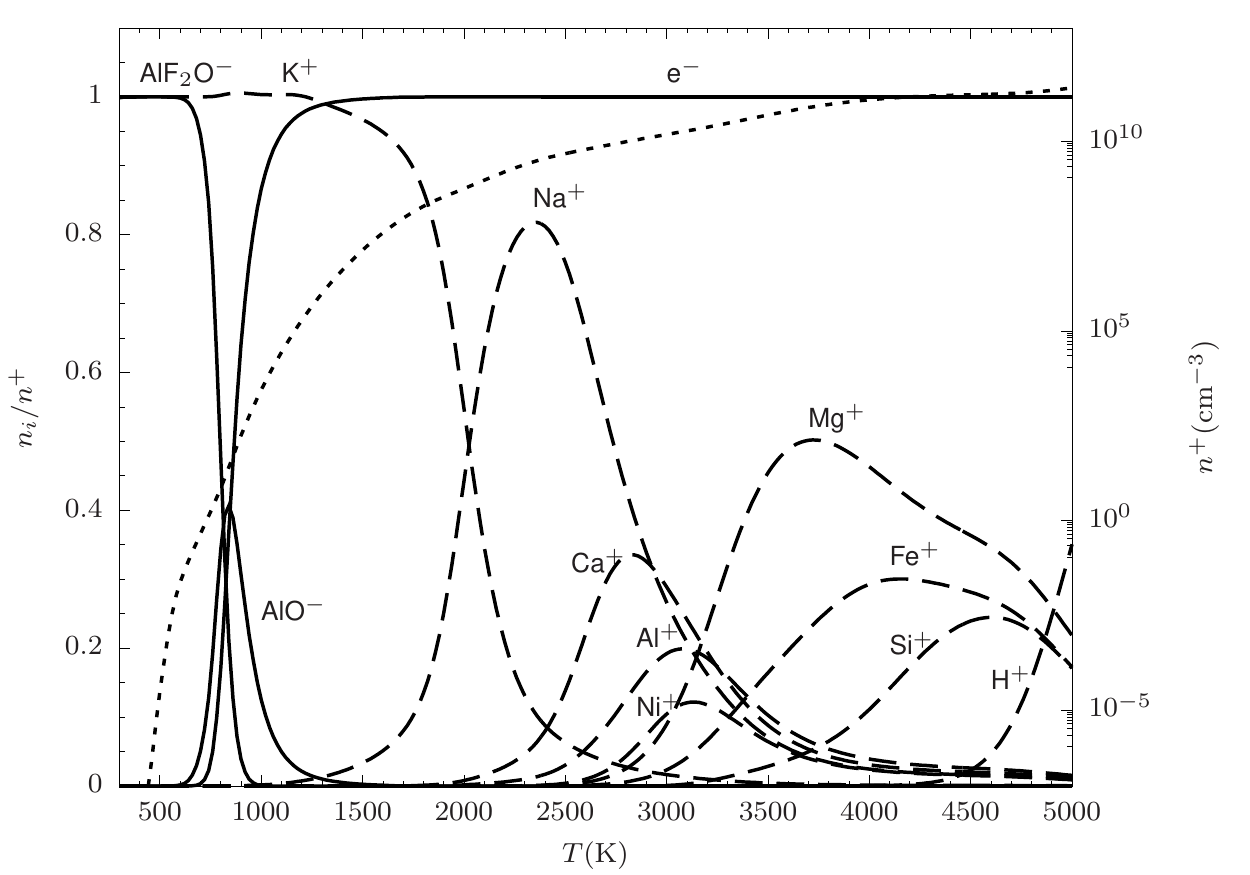}} 
	\caption{Electron number density and number densities of ions relative to the total number density of cations (dotted line). Electron and anions are marked with solid lines, cations with dashed lines. } 
	\label{fig:ions} 
\end{figure}

Figure~\ref{fig:sh6} shows the partial pressures of all atomic species in our model.
Besides the expected deviations due to the updated element abundances $\epsilon_j$ used here (cf. Table~\ref{table:elementabundances}), we find a decrease in the partial pressure of potassium (K) with decreasing reciprocal temperature ($\theta < 3\,\mathrm{K}^{-1}$).
The total number density of positively charged ions
\begin{equation}
n^+ = \sum_{\underset{\nu_{i0}<0}{i\in \mathcal{S}\setminus\mathcal{E}}}n_i
\end{equation}
as well as the electron density $n_0$ increase with increasing temperature (Fig.~\ref{fig:ions}, dotted line).
Above $T\approx800\,\mathrm{K}$ the charge balance is determined by free electrons and monatomic cations.
The maxima of the positively charged ions are distributed according to the ionization energies of the corresponding elements with the exception of Al and Ca.
The exception is due to the temperature dependence of the partition functions.
Below $T\approx800\,\mathrm{K}$ free electrons become very rare and the charge balance is primarily determined by K$^+$ and the polyatomic anions AlO$^-$ and AlF$_2$O$^-$ in our set of species shown in (Table~\ref{table:species}).
Note, that the $n^+$ already dropped to $1\,\mathrm{cm^{-3}}$.
The low ionization energy of potassium (4.3406633\,eV, \citep{Sug85}) leads to its ionization at relatively low temperatures and a decrease in the number density of potassium with decreasing reciprocal temperature $\theta$ (see Fig.~\ref{fig:sh6}).

\subsection{Convergence behavior}
\begin{figure}
	\resizebox{\hsize}{!}{\includegraphics{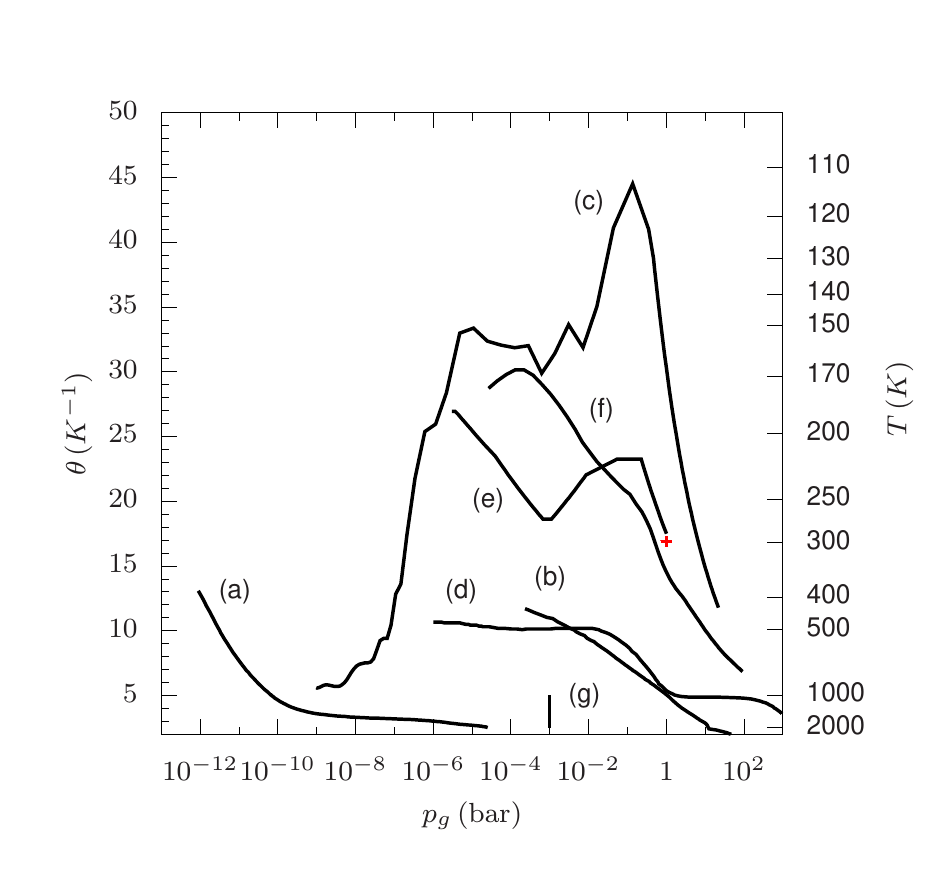}} 
	\caption{Selected thermal structures. The red cross marks the reference state for thermochemical data. (a) Circumstellar shell of an AGB star \citep{Gai87}, (b) Brown dwarf star GJ229b \citep{Tsu99} (c) Jupiter's atmosphere from measurements by the Galileo probe \citep{Sei98} (d) super-Earth GJ1214b with solar abundances \citep{Mil10}, (e) Earth's atmosphere up to $86\,\mathrm{km}$ altitude \citep{Coe76} (f) Venus reference atmosphere \citep{Sei85} (g) ($p_\mathrm{g}$,$\theta$) data used by \citet{Sha90} and in Figs.~\ref{fig:sh1} and \ref{fig:sh6}.} 
	\label{fig:profiles} 
\end{figure}

In this section, we study the efficiency of \texttt{FastChem} by simply determining how many iterations \texttt{FastChem} requires to meet the convergence criterion Eq.~(\ref{eq:convergence}) with $\delta=10^{-6}$ for selected test scenarios.
For this numerical test we use the same set of species as in the previous section (see Table~\ref{table:species}) but vary the gas pressures $p_\mathrm{g}$ between $10^{-12}\,\mathrm{bar}$ and $10^3\,\mathrm{bar}$ and the reciprocal temperature between $2\,\mathrm{K^{-1}}$ and $50\,\mathrm{K^{-1}}$ yielding a grid of 49\,600 points.
Note, that these extreme parameter ranges are selected exclusively for numerical tests.
Typical $(p_\mathrm{g},\theta)$ profiles of some astrophysical objects are shown in Fig.~\ref{fig:profiles}.
These profiles are for illustration only.
An in-depth interpretation with respect to these objects is not recommended since some of these atmospheres are clearly not in chemical equilibrium.

Therefore, \texttt{FastChem} was called for each of the 49\,600 grid points separately.
For a typical $(p_\mathrm{g},\theta)$ combination such as $p_\mathrm{g}=10^{-5}\,\mathrm{bar}$ and $\theta=15\,\mathrm{K}^{-1}$, 6 iterations in the oxygen-rich case are needed to fulfill the convergence criterion.
Usual computation times for \texttt{FastChem} are in the low milliseconds range for 510 species, including ions, on a standard desktop computer. 
For smaller systems restricted to neutral and ionized species composed of C, H, O, N, e$^-$ execution times can be reduced to microseconds.
Without electrons, the computational time can be considerably smaller.
The computational effort can increase to seconds for very low temperatures combined with high pressures.

\begin{figure}
\resizebox{\hsize}{!}{\includegraphics[trim = 2cm 0cm 0cm 0cm]{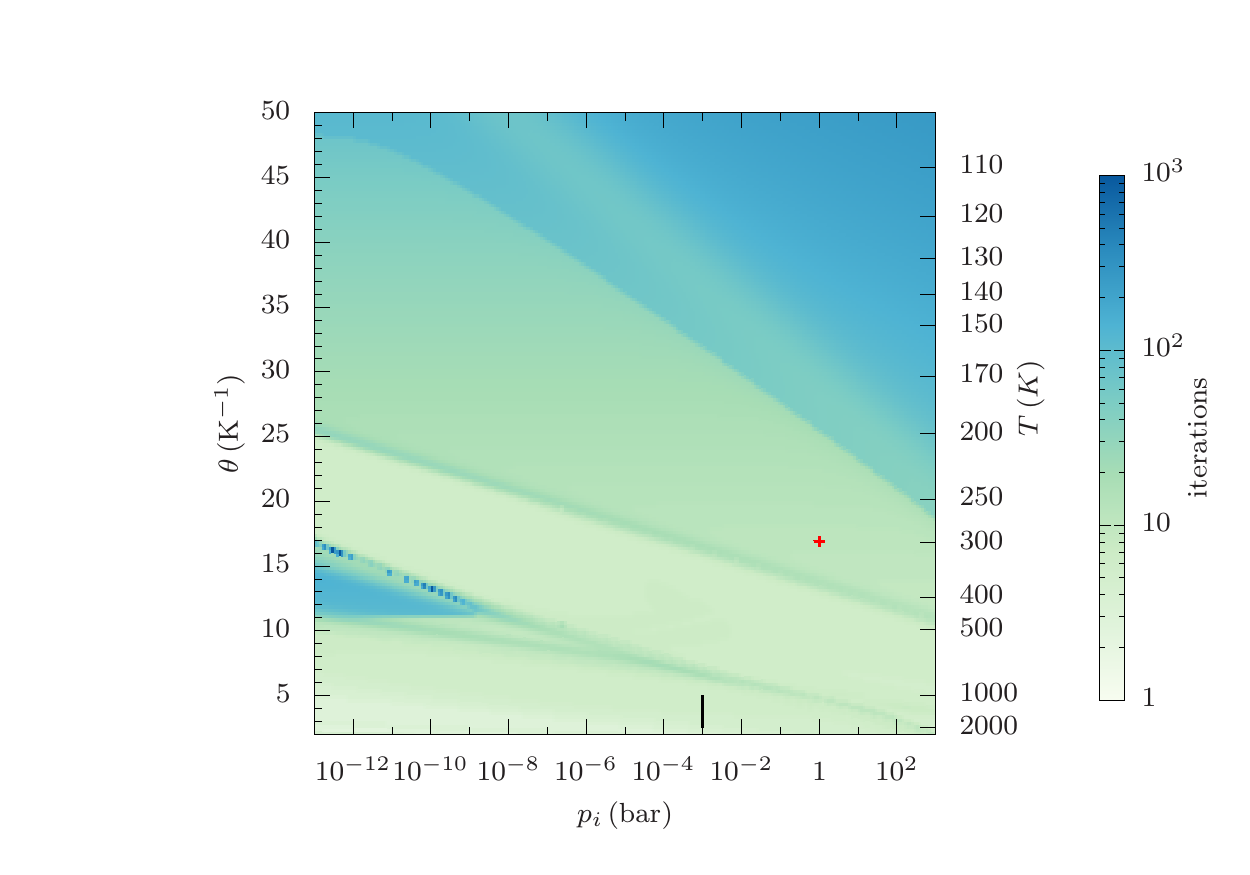}} 
\resizebox{\hsize}{!}{\includegraphics[trim = 2cm 0cm 0cm 0cm]{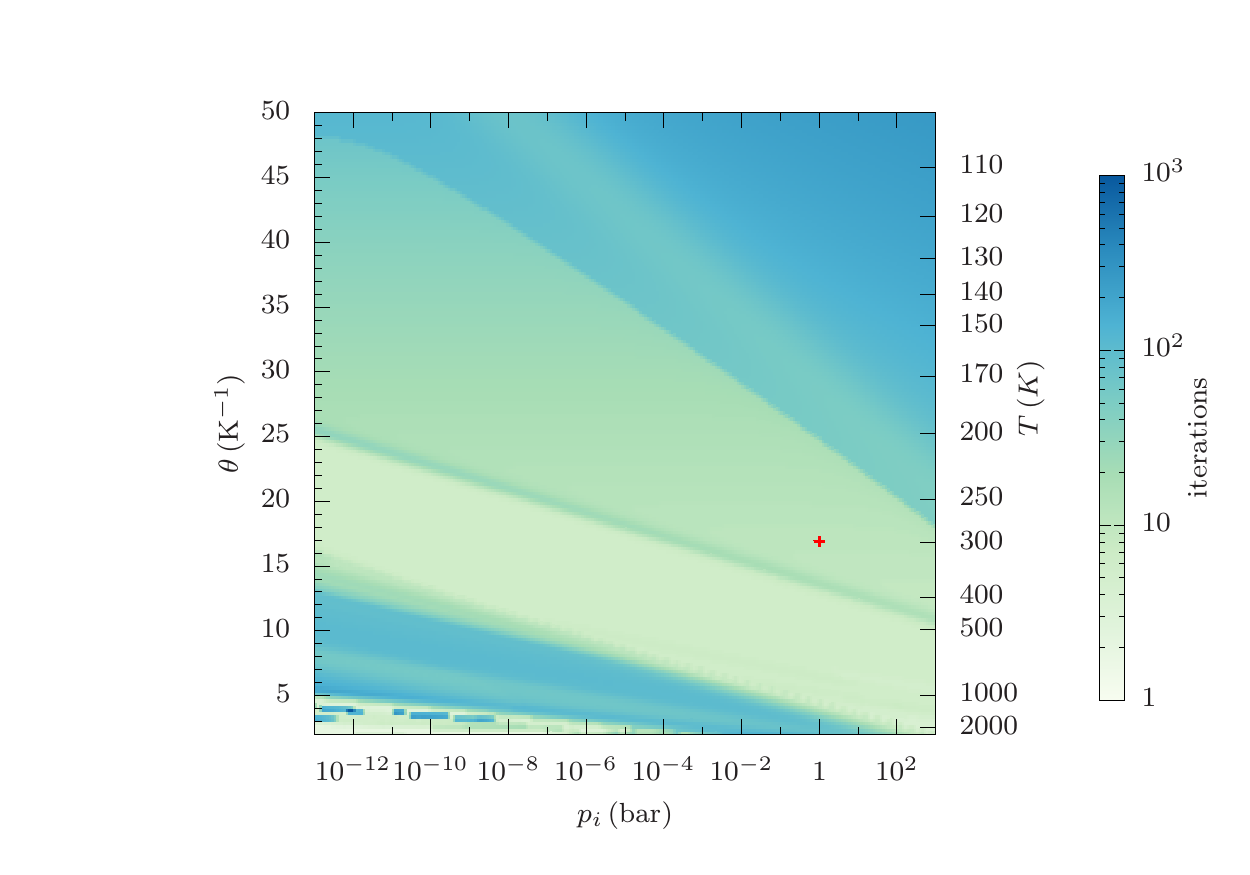}} 
\resizebox{\hsize}{!}{\includegraphics[trim = 2cm 0cm 0cm 0cm]{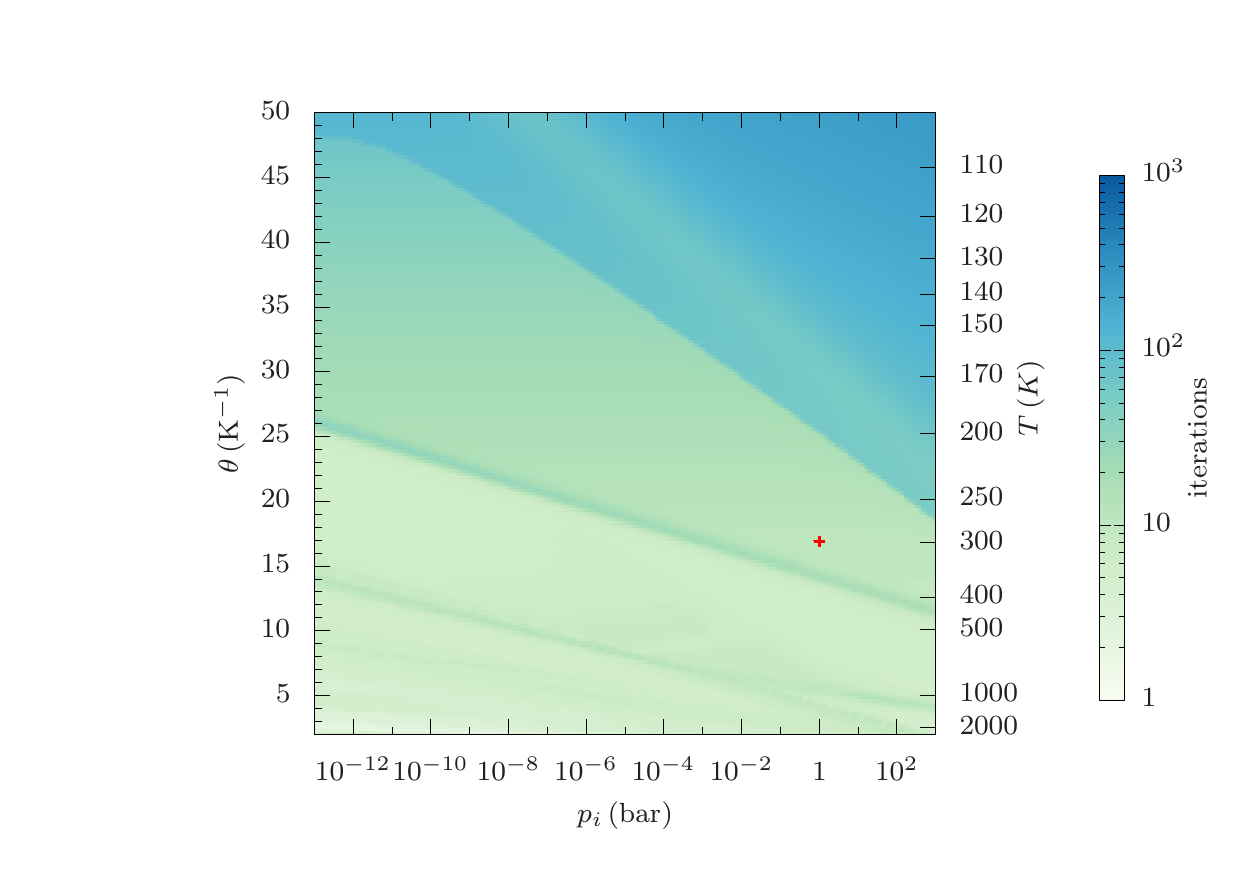}} 
\caption{Total number of iterations of the chemistry at the last pressure iteration step. The C:O ratio is (C:O)$_\mathrm{solar}$ (upper panel), 1 (mid panel) and 1/(C:O)$_\mathrm{solar}$ (lower panel).
The red cross marks the reference state for thermochemical data. ($p_\mathrm{g}$,$\theta$) data used in Figs.~\ref{fig:sh1} and \ref{fig:sh6} are marked by a black solid line in the upper panel.} 
\label{fig:convergence} 
\end{figure}

Three different chemical scenarios are studied in the specified $(p_\mathrm{g},\theta)$-plane:
\begin{itemize}
	\item oxygen-rich (element abundances according to Table~\ref{table:elementabundances})
	\item carbon abundance equals oxygen abundance (i.e. $\mathrm{C:O} = 1$, otherwise Table~\ref{table:elementabundances})
	\item carbon-rich (Table~\ref{table:elementabundances}, but with the exchanged values of $x_\mathrm{C}$ and $x_\mathrm{O}$)
\end{itemize}
Figure~\ref{fig:convergence} shows the total number of iterations of the chemistry at the last pressure iteration step as function of $p_\mathrm{g}$ and $\theta$.
Convergence is reached for all points in the $(p_\mathrm{g},\theta)$ plane shown.
In general the number of iterations increases with increasing reciprocal temperature and total gas pressure.
However, this behavior is not monotonously.
There are some features, where the total number of iterations becomes relatively large, most likely due to the competition between molecules for one element such as e.g. O.
The effect is especially pronounced in the cases where C:O$\lessapprox$1.

While the calculations presented in this work are only focussed on hydrogen-rich cases, \texttt{FastChem} has been also successfully tested for environments, where hydrogen is only a minor species.
However, the current version of \texttt{FastChem} requires a comparatively long calculation time for these cases because the pressure iteration is done via $n_\mathrm{<H>}$ (see Section~\ref{ssec:pressure}).
In a future version of \texttt{FastChem} we will adapt the code to remove the explicit dependence on hydrogen and replace it with the major element present.

\section{Summary}
For the efficient calculation of complex gas phase chemical equilibria, we present a semi-analytical, flexible computer program, called \texttt{FastChem}.
The program is written in object-oriented C++ which makes it easy to couple the code with other programs, although a stand-alone version is provided.
\texttt{FastChem} can be used in parallel or sequentially and is available under the GNU General Public License version 3 at \url{https://github.com/exoclime/FastChem} together with several sample applications.
The code has been successfully validated against previous studies and its convergence behavior has been tested even for extreme physical parameter ranges down to $100\,\mathrm{K}$ and up to $1000\,\mathrm{bar}$.
\texttt{FastChem} shows a stable and robust convergence behavior in even most demanding chemical situations (e.g. electrons competing with multiple anions at low temperatures and densities, equal carbon and oxygen element abundances, very large molecules consisting of more than 300 atoms), which posed sometimes to be extremely challenging for previous CE codes.

\section*{Acknowledgements}
DK aknowledges financial and administrative support by the Center for Space and Habitability and the PlanetS National Centre of Competence in Research (NCCR).



\bibliographystyle{mnras}
\bibliography{bibtex/references} 




\appendix
\section{Evaluation of products and scaling}
\label{sec:product}
Throughout our calculations, expressions of the form
\begin{equation}
p_{ij} = K_i \underset{l\neq j}{\prod_{l\in\mathcal{E}_0}} n_l^{\nu_{il}}\ ,
\label{eq:mwgmath}
\end{equation}
are ubiquitous in the algorithm.
The factors in Eq.~(\ref{eq:mwgmath}) can differ in hundreds of orders of magnitude. 
To evaluate these products and avoid numerical overflow  we use the equivalent expression
\begin{equation}
p_{ij} = \exp\left\lbrace\ln K_i + \underset{l\neq j}{\sum_{l\in\mathcal{E}_0}} \nu_{il} \ln n_l\right\rbrace \ .
\end{equation}
Additionally we define
\begin{equation}
\psi_j:=\max_{i\in\mathcal{S}\setminus\mathcal{E}}\left(\ln K_i + \underset{l\neq j}{\sum_{l\in\mathcal{E}_0}} \nu_{il} \ln n_l\right) - \xi_j\ ,
\end{equation}
where $\xi_j\geq0$ are constants.
Since the solution of the equation
\begin{equation}
P_j(n_j)=0
\end{equation}
is invariant against multiplication with a scaling factor $e^{\psi{_j}}$, we can write
\begin{equation}
\hat{P}_j(n_j)=\sum_{k=0}^{N_j}\hat{A}_{jk}n_j^k=e^{\psi{_j}}P_j(n_j)=0
\end{equation}
with the coefficients
\begin{align}
\hat{A}_{j0}&=e^{-\psi_j}A_{j0}=e^{\ln n_{j,\mathrm{min}}-\psi_j} - e^{\ln\left(\epsilon_j n_{<\mathrm{H}>}\right) -\psi_j}\ ,\\
\hat{A}_{j1}&=e^{-\psi_j}A_{j1}=e^{-\psi_j}+\underset{\epsilon_i=\epsilon_j}{\underset{\nu_{i j}=1}{\sum_{i\in \mathcal{S}\setminus\mathcal{E}}}} \exp\left\lbrace \ln K_i + \underset{l\neq j}{\sum_{l \in \mathcal{E}}} \nu_{i l}\ln n_l-\psi_j\right\rbrace\ ,\\ 
\hat{A}_{jk}&=e^{-\psi_j}A_{jk}=k\underset{\epsilon_i=\epsilon_j}{\underset{\nu_{i j}=k}{\sum_{i\in \mathcal{S}\setminus\mathcal{E}}}} \exp\left\lbrace\ln K_i + \underset{l\neq j}{\sum_{l \in \mathcal{E}}} \nu_{i l}\ln n_l - \psi_j\right\rbrace\ .
\end{align}

\section{Analytic expression of the electron density}
\label{sec:edensity}
Consider the case $\left|\nu_{i0} \right|\leq1$.
Then, in analogy to equation Eq.~(\ref{eq:basiceqalt}), we write
\begin{equation}
0 = n_0\left(1+\underset{\nu_{i0}=-1}{\sum_{i\in\mathcal{S}\setminus\mathcal{E}}}K_i\prod_{j\in\mathcal{E}}n_j^{\nu_{ij}}\right)-\frac{1}{n_0}\underset{\nu_{i0}=1}{\sum_{i\in\mathcal{S}\setminus\mathcal{E}}}K_i\prod_{j\in\mathcal{E}} n_j^{\nu_{ij}}
\end{equation}
This equation can be solved for $n_0$ directly and it follows
\begin{equation}
n_0=\sqrt{\frac{\alpha}{\beta}}\ .
\end{equation}
with
\begin{equation}
\alpha=\underset{\nu_{i0}=1}{\sum_{i\in\mathcal{S}\setminus\mathcal{E}}}K_i\prod_{j\in\mathcal{E}} n_j^{\nu_{ij}}
\end{equation}
and
\begin{equation}
\beta=1+\underset{\nu_{i0}=-1}{\sum_{i\in\mathcal{S}\setminus\mathcal{E}}}K_i\prod_{j\in\mathcal{E}}n_j^{\nu_{ij}}\ .
\end{equation}


\bsp	
\label{lastpage}
\end{document}